\documentclass{an}
\usepackage{times}
\usepackage{graphicx}
\graphicspath{{./fig/}{./png/}}
\usepackage{fancyhdr}
\usepackage{url}
\usepackage{bm}
\newcommand{\emf}{{\vec{\cal E}}}
\newcommand{\yr}{\,{\rm yr}}

\begin{document}

\title{Modeling a Maunder Minimum}
\author{A. Brandenburg\inst{1} \and E. A. Spiegel\inst{2}}
\institute{
NORDITA, Roslagstullsbacken 23, SE-10691 Stockholm, Sweden
\and
Astronomy Department, Columbia University, New York 10027, USA
}

\abstract{
We introduce on/off intermittency into a mean field
dynamo model by imposing stochastic fluctuations
in either the alpha effect or through the inclusion of a fluctuating
electromotive force.  Sufficiently strong small scale fluctuations 
with time scales of the order of 0.3-3 years can produce long 
term variations in the system on time scales of the order of hundreds of years.
However, global suppression of magnetic activity in both
hemispheres at once was not observed.
The variation of the magnetic field does not resemble that of the 
sunspot number, but is more reminiscent of the $^{10}$Be record. 
The interpretation of our results focuses
attention on the connection between the level of magnetic
activity and the sunspot number, an issue that must be
elucidated if long term solar effects are to be well understood.
}

\maketitle

\section{Introduction}

The sun shows variability on a broad range of time scales,
from milliseconds to millennia and on to even  longer scales.
Here we are interested in the time scales associated with
the solar cycle and its grand minima.
A salient manifestation of the cyclic behavior is seen in the
sunspot number, whose annual mean oscillates on a scale of eleven
years (or twenty-two years, if one goes by magnetic
polarity variations). The term ``cycle'' is used to describe this
oscillation in the sense that the sunspot number qualitatively 
performs the same kind of
oscillation approximately every eleven years, with an
amplitude that varies  in a way that is reminiscent
of some chaotic oscillators.  However, it has been found that 
we do not have sufficient data to decide whether  the global 
solar magnetic variation is  chaotic, in the sense that its temporal 
behavior may be that of a low-order deterministic dynamical 
system (Spiegel \& Wolf 1987).   Nevertheless, a chaotic oscillator
offers a natural way to model various irregularities of the solar
cycle (Spiegel 1977; Tavakol 1978; Ruzmaikin 1981).

Like certain simple chaotic systems, the sun repeats itself
magnetically over and over again, but never quite the same 
way twice, much in the manner of simple chaotic oscillators.
On the other hand, the simplest chaotic dynamos (Allan 1962; 
Robbins 1979) do not exhibit the kind of strong intermittency such
as was seen in the Maunder Minimum (Eddy 1978) when
the amplitude of the oscillation in sunspot numbers went
nearly to zero for seventy-five years in the time of Newton.
This behavior is suggestive of the possibility that a
strongly intermittent chaotic oscillator needs to be invoked
in trying to understand the solar oscillation. Oscillators
that behave this way are known (Spiegel 1981; Fautrelle \& Childress 1982) 
but, when they are in the active phase, their variations do not normally
resemble those of the sunspot number.  

In modeling the solar magnetic variability with simple oscillators,
it is not obvious how to connect the output of the models with the
sunspot number. It may therefore not be damning if the
variations produced by a model do not reproduce even
qualitatively the variations seen in the sunspot number.
Nevertheless, in terms of nonlinear lumped models, whose behavior is
purely temporal, it has been possible to produce
variations of the cyclic behavior that resemble those seen in the
grand minima qualitatively either by modulation (Weiss et al.\ 1984)
or intermittency (Platt et al.\ 1993b). But the solar
variability is manifestly spatio-temporal and lumped
models can at best provide clues to the actual processes involved in
the solar cycle. Here we describe an attempt to
go beyond models with purely temporal variations to 
models showing spatio-temporal variations that also produce analogues
of the grand solar minima.

An earlier version of this paper was composed in the early nineties and
some added remarks were inspired by discussions at the Enrico Fermi School
in Varenna (Cini Castagnoli \& Provenzale 1997).
In the meantime a lot of new work has emerged, but we feel that the
ideas presented here are still relevant.
Particularly important has been the work of Beer et al.\ (1998) in
demonstrating the persistence of a solar activity cycle throughout
the time of the Maunder Minimum.
Such behavior emerged from a purely temporal model 
(Pasquero 1996) as well as from a one-dimensional dynamo model (Tobias 1996)
with small turbulent magnetic Prandtl number (so that the viscous
diffusion timescale is much longer than that for magnetic diffusion time).
Similar results --- also for small turbulent magnetic Prandtl numbers ---
have been obtained for two-dimensional models by incorporating
quenching in the $\Lambda$-effect
that drives the differential rotation (K\"uker et al.\ 1999).
Intermittent behavior in dynamo models has been studied further by
Tworkowski et al.\ (1998) using time-dependent alpha-quenching, and by
Charbonneau et al.\ (2005) using algebraic alpha-quenching.
However, intermittency can equally well occur in stochastically
forced models (e.g.\ John et al.\ 2002).  In the following, we discuss the 
spatio-temporal variability of similarly forced dynamo models in two 
dimensions assuming axisymmetry.

\section{Spatio-temporal variability}

The spatio-temporal dynamics of the solar cycle as seen in
spacetime diagrams like the Maunder butterfly diagram
suggest that solitary waves may play an active part in the solar
activity cycle (Proctor \& Spiegel 1991).  A nonlinear version of
Parker's (1955) dynamo waves (Worledge et al.\ 1997)
or solitary waves arising from another overstability
could be the mechanism of the drifting
of the center of activity in latitude through the
cycle. The finite width of the activity zone at any given time
suggests that the waves in question are themselves confined or
guided by a layer of some corresponding thickness.
The picture that we adopt here is the now conventional
one that the layer is the tachocline, though such a layer
has been variously considered to lie deep in the convection
zone (DeLuca 1986), just below it (Spiegel \& Weiss 1980) or
occupy the full convective zone (Brandenburg 2005).
It might operate on
the standard ingredients of a dynamo --- differential
rotation and cyclonic convection --- from which we here
make a model.  Since we first wrote those lines, the
layer has been well studied both observationally and theoretically 
(Hughes et al.\ 2006).  It mediates the transition between the outer 
differentially rotating layers of the convection zone 
and the inner core with its nearly constant angular velocity.  
It has been renamed  the tachocline (Spiegel \& Zahn 1992)
in keeping with its gain in  respectability.
  
Here we use mean field theory (Moffatt 1978; Krause \&
R\"adler 1980) with the effects of small scale motions
subsumed into turbulent diffusivity and the $\alpha$-effect. Though
mean field dynamos may not tell the whole story of the
solar magnetic fluctuations (e.g.\ Hoyng 1987; 1988), they
will suit our purpose of modeling the intermittency signaled by 
the Maunder Minimum.   In modeling solar intermittency, we need to 
be aware that there are several forms of intermittency that have 
been isolated in dynamical systems theory. However, there is 
a particular one that models the solar grand minima quite well 
(Pasquero 1996) and that has come to be called on/off intermittency 
(Platt et al.\ 1993a; see also Spiegel 1994).  [In fact,  this form of 
intermittency was fashioned (Spiegel 1981) with grand minima in mind.]

On/off  intermittency is like the intermittency detected
in the output of a probe in a turbulent fluid registering
abrupt changes  as laminar and turbulent fluid regions flow past it.
One may think of this as a series of bursts or
chaotic relaxation oscillations. What characterizes
models that have been made of this process (e.g.\ Spiegel 1981;
Ott \& Chen 1990; Pikovsky \& Grassberger 1991) is that
the (potentially) unstable oscillator performing the cycle
is driven to instability through coupling to an aperiodic
driver that is continuously chaotic or stochastic. The role
of the driver is to move the system into and out of the
unstable state. Observations of the bursts cannot readily
distinguish the chaotic driver from a stochastic
alternative which works equally well (von Hardenberg et al.\ 1997),
and the number of degrees of freedom involved in such an
object in the solar case will be hard to determine (but see
Heagy et al.\ 1994). Fortunately, the qualitative features of
the process do not depend sensitively on this difference,
and we shall use a stochastic driver here.  In our present 
considerations, the action of the driver is meant to 
model the influence of the convective 
solar dynamo and the on/off oscillator represents the tachocline.
Here we introduce on/off intermittency into a working model of the solar 
cycle (R\"udiger \& Brandenburg 1995) that has not previously 
produced grand minima.   Modulational grand minima in dynamo
models have been found by Tobias.
In the case of a distributed dynamo (Brandenburg 2005), on/off
behavior may similarly be produced.

The effects of stochastic noise on mean field
$\alpha\Omega$ dynamos have been studied previously (Choudhuri 1992;
Moss et al.\ 1992;  Hoyng et al.\ 1994), mainly to model
irregularities of the solar cycle on time scales comparable
to the solar cycle itself or shorter. Moss et al.\ (1992)
suggested that a modulation of the solar cycle on longer
timescales of the order of centuries should also be possible.
More recently, work along those lines (Schmitt et al.\ 1996)
has introduced the on/off intermittency mechanism into a
mean-field style of dynamo, as had been used already in a
lumped model of the solar cycle (Platt et al.\ 1993b). The
present paper is similarly based on a relatively realistic
model of the solar dynamo in being spatio-temporal and 
into  which we introduce the on/off intermittency mechanism.

The main difference between
our work and that of Schmitt et al.\ is that we
consider a standard $\alpha$-effect while they
introduced a lower cutoff excluding field generation below a
field strength of 1 kG at the base of the convection zone.
This kind of filtering is analogous to
Durney's (1995) introduction of a critical field value in the
tachocline above which a magnetic tube is ejected. This also
produces grand minima in spatio-temporal models.   

Another difference with the calculation of Schmitt et al.\
is that we consider the full sun whereas they studied only
a single hemisphere and produced one-winged butterflies.
This relates to another aspect of the problem that our
model is intended to bring out.  It is very difficult to have
both hemispheres go completely inactive for appreciable times.
The probability of a complete turnoff is exceedingly low and it
seems unlikely that any propagative intermittency mechanism 
with retardation effects could turn off the magnetic activity globally
in models with large spatial extent.   The model described here
does however produce lowered solar activity over large portions
of its computational domain and therefore gives the kind of 
reduction in total overall activity that is consistent with what was 
seen in the Maunder Minimum.

The hemispheric asymmetry has been modeled by Knobloch et al.\ (1998;
see also Weiss 1993) and this work is related to the problem of making
an extended system demonstrate on/off intermittency.  The problem has
something in common with that involved in laying a rug.
Once the rug is nailed down, if there is a bump somewhere,
there seems to be no way to squeeze it out of existence. We
have observed the same behavior with the dynamo model. If we
produce a grand minimum locally somewhere in space, we
find invariably that there is usually some excitation elsewhere.   
While we can get a whole hemisphere to be quiet at
one time, the other one may still show activity. The output of
the dynamo model is therefore somewhat like the sun in intermission ---
on the global scale there is usually some weak activity somewhere 
and it  does have cycles.  [Pasquero (1996) has shown how this may be achieved
in the purely temporal oscillators as well.]  As we have learned
at the Fermi School on the Interaction of the Solar Cycle
with Terrestrial Activity in June of 1996, certain terrestrial
data have much more in common with the output of
the dynamo models than the sunspot number. Moreover,
some of these terrestrial data are very convincing proxy
data for the solar activity (Beer et al.\ 1990; 1994; 1996;
Solanki et al.\ 2004), while others may mimic the sunspot number. 
[There are also auroral indicators of solar activity recorded in classical
antiquity (Stothers 1979, Solow 2005).]   To a great extent, the issue is
at heart  one of knowing how to compare the output of dynamo models
to the various measures of solar activity. 

We describe next some particulars of the model itself,
including the manner in which the fluctuations
are introduced. Readers who are not interested in such
detailed information should skip directly to Sect.~4 where
we outline the main results. In these, we focus on models
with a rather high noise level exceeding the electromotive
force from the $\alpha$-effect by an order of magnitude since
we expect the global convective dynamo action to be more 
vigorous than that of the tachocline.

\section{The model}

We use a mean-field model of the dynamo action in the
tachocline (R\"udiger \& Brandenburg 1995) with an anisotropic 
$\alpha$-effect and a turbulent magnetic diffusivity.   Magnetic
buoyancy is also included as it is in the more elaborate model of
Jiang et al.\ (2007; see further references therein). 
The prescribed angular velocity is taken from
the observational findings of helioseismology (Christensen-Dalsgaard 
\& Schou 1988). To obtain a 22 yr magnetic cycle period, we introduce a
scaling factor in the magnetic diffusivity of 0.5 and, to get a
butterfly diagram with sufficient activity at low latitudes,
we introduce a suitable latitude dependence in $\alpha$.

In the original model calculations, the value of $\alpha$ was
typically set at a value approximately twenty times the
critical value for instability. To produce on/off intermittency
(Platt et al.\ 1993a), in the highly supercritical case,
we would need very large fluctuations at that value. In
this exploratory study, we prefer to operate at more
modest parameter values to avoid the need
for  large fluctuations in the driver.
Therefore we choose a slightly subcritical
value of $\alpha$ and introduce only modest fluctuations in its
magnitude so that the driver can move it into and out of
the unstable state easily.  For this purpose we introduce a scaling factor
$c_\alpha$ in front of certain components of the $\alpha$-tensor (those
components which result from the interaction of rotation
and stratification). For details see the original discussion
of the model (R\"udiger \& Brandenburg 1995) where, with
$c_\alpha$ = 1, the resulting toroidal magnetic field is a few kgauss
and the poloidal field at the surface about 10 gauss. In
the cases studied here, however, where the dynamo is just
marginally excited, the generated magnetic field is weak,
and could not explain the field strength observed in
sunspots.  This feature of the model can be corrected, as the
work of Schmitt et al.\ (1996) shows.

We consider separately two kinds of fluctuation in the
model:  fluctuations in $\alpha$ or in an imposed
electromotive force $\emf$ representing the effect
of the fluctuations in the main solar dynamo. The latter is 
a more elementary process that does not rely much on dynamo
theory. The picture here is that in the bulk of the convection
zone a small scale dynamo operates (e.g.\ Meneguzzi
\& Pouquet 1989; Nordlund et al.\ 1992) producing a magnetic 
field that is highly variable in space and time as
represented by $\emf$.  In the bulk of the convection zone the
fluctuations are immense, hence even spatial and temporal
averages ($\emf=\langle\bm{u}'\times\bm{B}'\rangle$)
remain fluctuating, albeit on longer scales (Brandenburg et al.\ 2008).
The angular brackets refer
to ensemble averages in principle but, in practice, they are approximated
by spatial and temporal coarse graining averages.
The error resulting from this approximation is sometimes
interpreted as a source of stochastic noise (Hoyng 1987; 1988; 1993;
Moss et al.\ 1992; Brandenburg et al.\ 2008).  As we have mentioned, there is ample
reason for expecting fluctuations to appear in any realistic model.

We adopt white noise with vanishing mean value and a root 
mean square value of unity.
The temporal power spectrum of  the noise %$N = N(t, r, \theta)$, 
is flat for frequencies smaller than $2\pi/\tau$.
The value of $\tau$ determines the time span
over which long term variability of
the resulting mean magnetic field is possible.
The amplitude of the noise, $n_{\cal E}$, is measured in terms of the
rms velocity of the turbulent motions and the local equipartition
field strength. 
%%%  Axel, is this too subtle for the present discussion?  It leaves out penetration
%%% and so forth. 
%If the cycle-producing dynamo is in a
%stably stratified shear layer just below the convection zone  --- the
%tachocline --- the fluctuating emf should really act on the
%boundary only, but we have not attempted to implement
%this version of the forcing, which may be relevant to Parker's (1993)
%interface dynamo.  
%AB: OK, yes. So, I also removed then the Parker 93 from the reference list.
Whichever of the fluctuation mechanisms applies --- in the emf  or in the $\alpha$ effect
--- the procedure is similar.  For the $\alpha$ case, for example, 
we add a fraction $n_\alpha$ of the noisy component to the original $\alpha$.
 
\section{Results}

\subsection{Fluctuating $\alpha$-effect}
To regulate the strength of the effect of noise, we introduce a factor
$c_\alpha$ in front of the familiar $\alpha$ term, as described in
Sect.~3. With everything else fixed, instability occurs when $c_\alpha$
exceeds the critical value $c_\alpha^{\rm(crit)}\approx0.03$.
We begin by adopting the subcritical value $c_\alpha=0.02$ and
we adjust the fluctuations in $\alpha$ to have a time scale $\tau=3\yr$.

If the noise level, $n_\alpha$ , is too low, the magnetic field
decays to zero, as it does when $n_\alpha$ = 0.02. Short noisy bursts
in $\alpha$ are insufficient to bring the magnetic field to
appreciable strength. On the other hand, if $n_\alpha$ is rather larger
than this, the magnetic field is almost entirely dominated
by the noise and cyclic behavior does not occur, except
for irregular reversals on the timescale of centuries. Such
a case is depicted in Fig.~1, where we show a color-coded representation
of the toroidal magnetic field at the bottom of the convection zone
as a function of time and latitude.
In the following we refer to such representations as butterfly diagrams.
A very similar result, with reversals on a
long time scale, is found even when the non-random component
of $\alpha$ is absent altogether provided there is global
shear in the tachocline, as there is in accretion disks   (Vishniac \&
Brandenburg 1997).

To get sufficiently large magnetic field strengths, we
focus attention on models with $c_\alpha=1$. With $\tau=3\yr$ we
find long term variability on a time scale of $200$--$500\yr$; see
Fig.~2.    We associate this variability with grand minima (and maxima)
as discussed below in Sect.~5 although as noted already, there is 
never a complete turning off of the cycle at all locations at once.
 
Even if $\tau$ is decreased to a value of $0.3\yr$, say,
long term variability still occurs, but individual cycles may
vary significantly in amplitude as seen in Fig.~3. Such behavior
is found only if the fluctuations of $\alpha$ are sufficiently larger
than the average value; here a factor of five is needed.
Individual fluctuations of $\alpha$ can still be much larger because
we have assumed an exponential distribution for the probability
density of $\alpha$ fluctuations.

\begin{figure}[t!]\begin{center}
\includegraphics[width=\columnwidth]{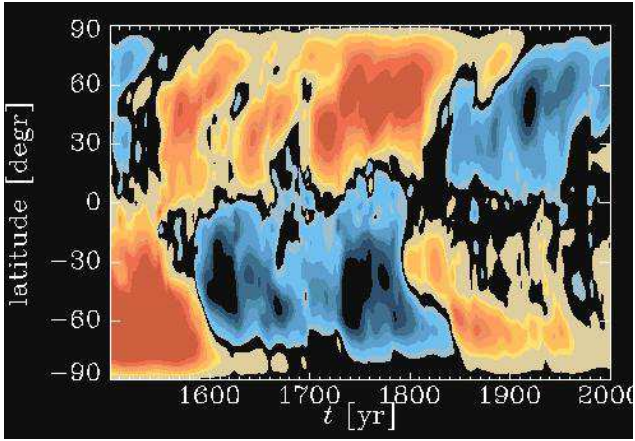}
\end{center}\caption[]{
Butterfly diagrams of the toroidal field at the base of
the convection zone for $c_\alpha=0.02$, $n_\alpha=5$ and $\tau=3\yr$.
}\label{fig1}\end{figure}

\begin{figure*}[t!]\begin{center}
\includegraphics[width=\textwidth]{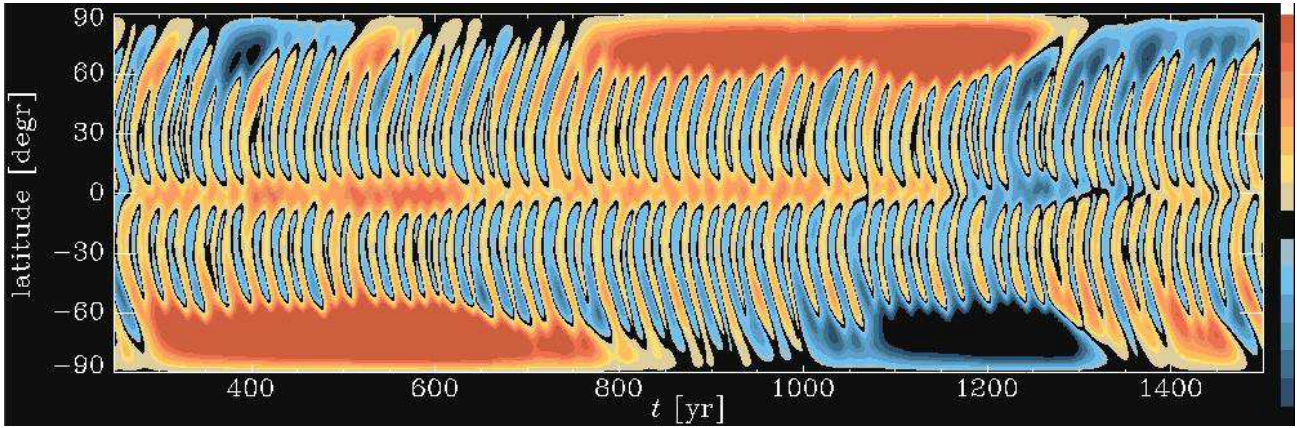}
\end{center}\caption[]{
Butterfly diagrams of the toroidal field at the base of
the convection zone for $c_\alpha=0.02$, $n_\alpha=5$ and $\tau=3\yr$.
Butterfly diagrams of the toroidal field at the base
of the convection zone for $c_\alpha$ = 1, $n_\alpha=0.2$ and $\tau=3\yr$.
}\label{fig2}\end{figure*}

\subsection{Magnetic noise}

If we introduce an external stochastic emf, as described in
Sect.~3, we find a qualitatively similar behavior to that
with fluctuating $\alpha$; see Fig.~4. Fluctuations of various
kinds can evidently produce long time intermittency.

The model shows two distinct activity waves, one
migrating equatorward and the other poleward. The two
waves seem to be modulated independently, in each hemisphere.
So do the modulations in the two hemispheres
seem to be only weakly coupled, with little tendency to
form a dipole structure. It may be that the approximate
antisymmetry of the toroidal magnetic field of the sun,
suggested by Hale's polarity law, may not be a very stable
feature, and that other types of (a)symmetry might have
occurred in the past. Other, more regular parity variations
of the magnetic field have previously been seen in nonlinear
models (Brandenburg et al.\ 1989a,b; 1990; Jennings \&
Weiss 1991; Sokoloff \& Nesme-Ribes 1994).

\section{Interpretation}

In the model described here, the seat of the solar activity cycle is in the solar tachocline, 
the layer that matches the differential rotation of the solar convection zone to the (nearly) 
rigid rotation of the inner sun (Hughes et al.\ 2006).   While the precise hydromagnetic
process that drives the activity has not yet been securely identified, we have assumed 
that the process may be modeled as an $\alpha\omega$ dynamo for the purpose of
exploring the cause of the grand minima of solar activity.    However, as we have already 
implied, any of several overstabilities might equally serve our purposes.   

As we have tacitly assumed,  the differential rotation in the tachocline is likely to be significant
in such instabilities.  In particular it is likely to  give rise to a toroidal field.  Given a suitable depth
dependence of the strength of this field, an instability driven by magnetic buoyancy (Parker 1979)
may arise and, especially in the presence of a stable density stratification, it may drive waves
of excitation (Proctor \& Spiegel 1991).   Another possible driver for such waves may be 
magnetorotational instability (Balbus \& Hawley 1998).  Parfrey \& Menou 
(2007) have studied the local instability  of the tachocline and found 
instability at high latitudes and stability at low latitudes.   However, their calculation did 
not include a toroidal field and, as Knobloch (1992) has observed, an azimuthal field can 
cause a Hopf bifurcation in MRI.   As we may reasonably expect to find a toroidal field
in the tachocline, this overstability is another mechanism that could perhaps engender
waves whose description would typically be by the complex Ginzburg-Landau equation
(Aranson \& Kramer 2002).  A related calculation is described by
Kitchatinov and R\"udiger (2007); see also Cally (2003), Gilman et al.\ (2007),
R\"udiger \& Kitchatinov (2007), and Zahn et al.\ (2007).

\begin{figure}[t!]\begin{center}
\includegraphics[width=\columnwidth]{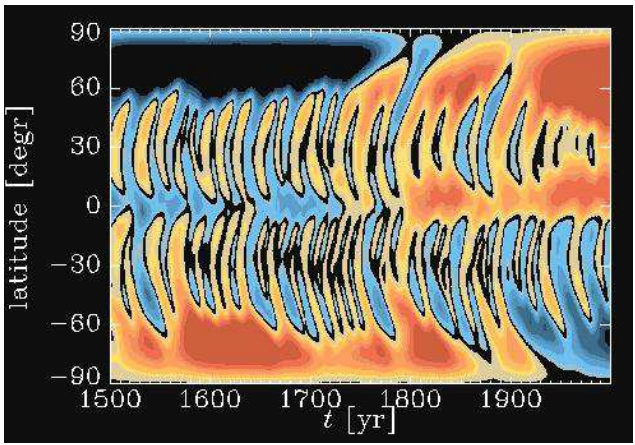}
\end{center}\caption[]{
Butterfly diagrams of the toroidal field at the base of
the convection zone for $c_\alpha=1$, $n_\alpha=5$ and $\tau=0.3\yr$.
}\label{fig3}\end{figure}

\begin{figure}[t!]\begin{center}
\includegraphics[width=\columnwidth]{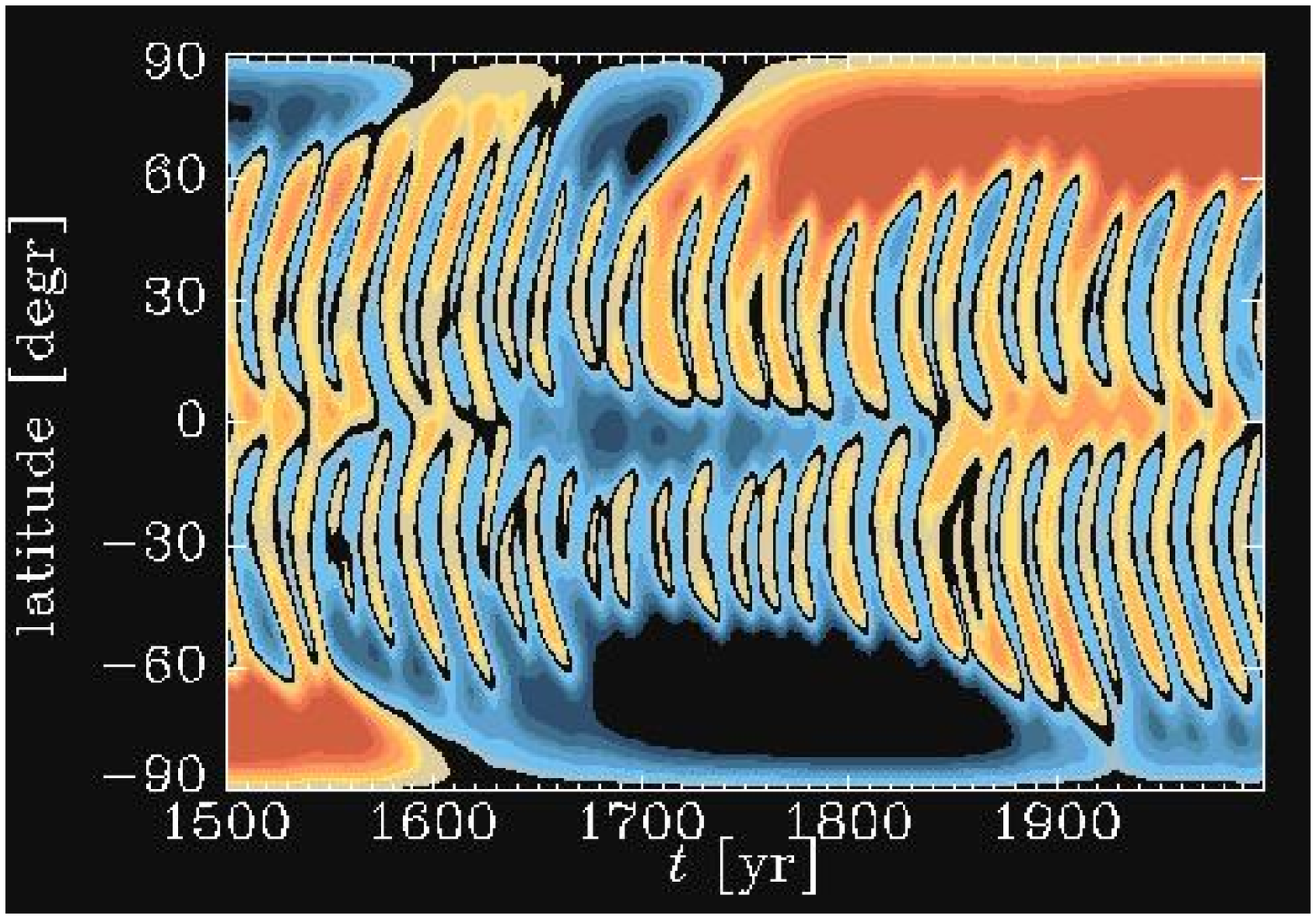}
\end{center}\caption[]{
Butterfly diagrams of the toroidal field at the base of
the convection zone for $c_\alpha=1$, $n_{\cal E}=1$ and $\tau=3\yr$.
}\label{fig4}\end{figure}

Given that the tachocline is poised to drive hydromagnetic activity, by dynamo processes or
otherwise, we have also assumed that the ambience is highly fluctuating.  Here we have in 
mind that the main dynamo in the sun is a global convective dynamo.   It is this process that
is presumably responsible for the magnetic carpet (Simon et al.\ 2001) in a continuous 
process that is thought to produce rapidly fluctuating fields of moderate strength 
(Priest et al.\ 2002).   In view of the 
complications in the theory of this process, we have simply modeled the highly fluctuating 
fields in of the solar convection zone as a stochastic process.    In the mechanism of
on/off intermittency that we have introduced here, the fluctuations produced by the main
dynamo move the tachocline into and out of states of instantaneous overstability.
This mechanism produces, as  we have noted, longer response times in the model tachocline 
than the time scales of the fluctuations that give rise to more intense field concentrations.
[We have seen a similar kind of symbiosis in the simulations of the geodynamo by
Glatzmaier \& Roberts (1995)].   The details of the process are complicated but, in gross, when 
the local field fluctuations produced are too weak, the grouping of flux ropes into a sunspot field 
is not achieved, even though much of the normal activity continues in the convection zone.   To
illustrate how such details may relate to the observed spot numbers, we have mapped the
activity to the butterfly diagram with a particular functional of the field in the tachocline for the
present purposes.
  
Like the sun, the model we have studied here manifests spatio-temporal
intermittency but the magnetic activity almost never turns off everywhere.
That is, our results suggest that simple dynamo models, with either fluctuating
emfs  and/or  with magnetic noise injected into the bulk of
the convection zone, can produce intermittency on
sufficiently long time scales, but they do not switch off globally
over the whole sun, in both hemispheres at once.
It seems likely that this feature is inherent in all models with
propagative behavior and that it cannot be expected that the
cycle-producing processes of the sun switches off everywhere
for several cycles as might be imagined on the basis of
certain lumped models (such as Platt et al.\ 1993b).
When we first produced the present results, we thought this feature
was a deficiency. But as Ribes \& Nesme-Ribes (1993) have reported, even
during the Maunder Minimum, a weak solar cycle continued.  
Thus a globally depleted activity level is more like what is wanted
and we have gotten that from the model in keeping with the historical 
records as interpreted by Ribes and Nesme-Ribes.

There still remains the need for an additional physical feature
that relates to the production of sunspots, or more specifically, strong flux tubes
of sufficiently large cross-section.   That is the message
of the procedures of Durney (1995) and Schmitt et al.\
(1996). A simple way of formulating this problem is to
say that the sunspot number, which is a global parameter,
is, as already mentioned, a functional of the various fields produced in the
model. To get that functional, we need to operate with an
explicit sunspot production mechanism.

Let $R(t)$ be the average magnetic field strength between
$\pm10^\circ$ and $\pm30^\circ$ latitude, but allow only those areas
where the field exceeds the root mean square value by 30
percent to contribute to this average. This quantity is
plotted in Fig.~5. Note that $R(t)$ shows periods with almost
vanishing magnetic activity as in a Maunder Minimum.
The grand minima on this interpretation of the output of
the models result from what may be regarded as lean cycles magnetically.
If there are magnetic droughts, they are
only local, not global, though the magnetic means may be
low. In that sense, the sunspot number, though valuable
for having focused our attention on an interesting feature
of the magnetodynamics, may in some ways be a misleading
indicator of what is happening overall.

\begin{figure}[t!]\begin{center}
\includegraphics[width=\columnwidth]{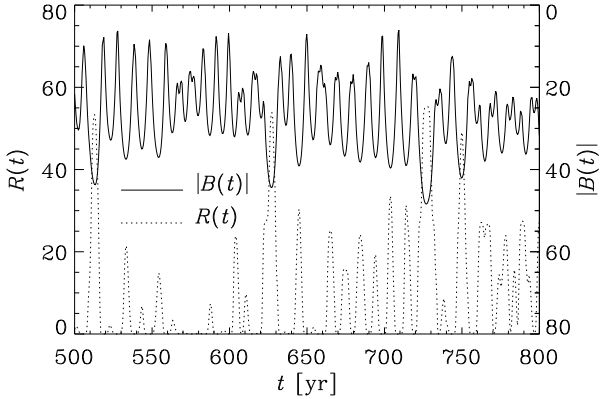}
\end{center}\caption[]{
Time series of $|B|$ (dotted line)
and the activity parameter $R$ (solid line)
for the same run as in Fig.~2. Note that
the scale for $|B|$
increases downwards, in order to mimic the
approximate anticorrelation between the $^{10}$Be data and the
sunspot number.
}\label{fig5}\end{figure}

This view is supported by studies of other indicators
of solar activity than the sunspot number. The most complete
records of variations that may result from solar activity
fluctuations are found in the $^{10}$Be records from ice
cores (Beer et al.\ 1990; 1994; 1996). The $^{10}$Be variations are
plausibly attributed to modulation of the cosmic ray flux
by the magnetic field in the solar wind (Beer et al.\ 1996).
In Fig.~6 we reproduce data kindly provided by Dr.\ J\"urg
Beer comparing the $^{10}$Be records with the sunspot number
for nearly four centuries. We see that a cyclic modulation
is very much in evidence during the Maunder Minimum,
with a rather modest reduction in its amplitude compared
to that of the sunspot number.

\begin{figure}[t!]\begin{center}
\includegraphics[width=\columnwidth]{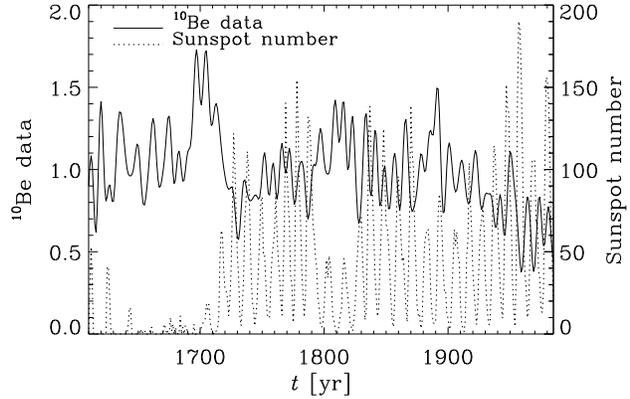}
\end{center}\caption[]{
$^{10}$Be data (solid line) together with the sunspot number
(dotted line), as provided by Dr.\ J\"urg Beer. The data are from
a shallow core (300 m) drilled at Dye 3, Greenland, in 1986.
The data were filtered using a spectral filter with a cut-off of 6
years and interpolated using a cubic spline. The younger part
(1783-1985) is published in (Beer et al.\ 1990), and the whole
record appeared in Beer et al.\ (1994).
}\label{fig6}\end{figure}

The situation as brought out in the cited papers of
Beer et al.\ is that various measures of solar activity are
not perfectly correlated. If we were to think of the output
of a model solar dynamo as we would one of these
other measures, we would not be surprised that it does
not necessarily represent all of them faithfully.
Unfortunately, there is as yet no clear theoretical indication
which one of them a model should most closely represent.
If, in a grand minimum, there is high magnetic activity in
only one hemisphere, it is not unreasonable that the
modulations of $^{10}$Be should continue with reasonable strength.
The question for the theory then is what is the relationship
between the variations produced by the models and (say)
the sunspot number. This is particularly difficult since the
sunspots seem to be produced well within the sun and not
at the surface, whereas some of the other activity
measures are no doubt superficially produced. Even if we do
have a good model of the cyclic mechanism, we also need
to understand how sunspots form and surface before we
can predict their number.

Nor is it unimportant to try to predict the sunspot
number, or something akin to it, for it is this quantity that
seems to be connected to some climatological variations.
The most striking evidence of this is the discovery in tree
ring data (Douglass 1927) that ``the sunspot curve flattens
out in a striking manner ... from 1670 or 1680 to 1727.''
This discovery was made by A.E. Douglas before he had
``received a letter from Professor E. Maunder ... calling
attention to the prolonged dearth of sunspots between 1645
and 1715.'' As Douglass and others have argued, variations
in tree ring thickness in turn are connected to rainfall, so
it is not an idle project to try to understand how to go
from the workings of a solar dynamo to the manufacture
of sunspots.

In summary, solar activity waves at high and low
latitudes and in the two hemispheres, lead somewhat
independent, weakly correlated lives and fluctuate separately
under the influence of noise. Spatial variations of the solar
cycle during the Maunder Minimum are not an immediate
indicator of the sunspot number and, if the sunspots
have their origin well beneath the solar surface, we must
go another step in the discussion before we obtain results
that are fully consistent with historic records of sunspots as
summarized by  Ribes and Nesme-Ribes (1993). It is not at 
all obvious what we may conclude about subconvective activity 
from observed surface activity.
 
\acknowledgements
We are grateful to Dr.\ J\"urg Beer for his
help and advice and for providing the data for Fig.~6. This
was made possible by our participation in the E. Fermi School
organized by Drs. J. Cini-Castagnoli and A. Provenzale. We
thank Dr. David Moss for making useful suggestions about the
manuscript and Dr. A. Solow for a discussion of his work.  A.B. is grateful 
for the hospitality of the Astronomy Department at Columbia University, 
where most of this work was carried out with support from the AFOSR under
under contract number F49620-92-J-0061.

%r e f

\end{document}